\documentclass[10pt,preprint2]{aastex}
\usepackage{epsfig}
\usepackage{amsmath}
\usepackage{amssymb}
\usepackage{ulem}

\newcommand{\ups}{\rule{0pt}{15pt}}
\newcommand{\asec}{${}^{\prime\prime}$}

\begin{document}

\title{Study on atmospheric optical turbulence above Mt.~Shatdzhatmaz in 2007--2013}

\author{V.~Kornilov, B.~Safonov, M.~Kornilov, N.~Shatsky, O.~Voziakova,  S.~Potanin,\\ I.~Gorbunov, V.~Senik, D.~Cheryasov}

\affil{Lomonosov Moscow State University, Sternberg Astronomical Institute\\
Universitetsky prosp. 13, 119992 Moscow, Russia}
\email{victor@sai.msu.ru}

\begin{abstract}
We present the results of the atmospheric optical turbulence (OT) measurements performed atop Mt.~Shatdzhatmaz at the installation site of new 2.5-m telescope of Sternberg Astronomical Institute. Nearly 300\,000 vertical OT profiles from the ground up to an altitude of 23~km were obtained in the period November 2007--June 2013 with the combined multi-aperture scintillation sensor (MASS) and differential image motion monitor (DIMM) instrument.

The medians of the main OT characteristics computed over the whole dataset are as follows: the integral seeing $\beta_0 = 0.96$~arcsec, the free-atmosphere seeing $\beta_{free} = 0.43$~arcsec, and the isoplanatic angle $\theta_0 = 2.07$~arcsec. The median atmospheric time constant is $\tau_0 = 6.57 \mbox{ ms}$. The revealed long-term variability of these parameters on scales of months and years implies the need to take it into account in astroclimatic campaign planning. For example, the annual variation in the monthly $\theta_0$ estimate amounts up to 30\% while the time constant $\tau_0$ changes by a factor of 2.5.

Evaluation of the potential of Mt.~Shatdzhatmaz in terms of high angular resolution observations indicates that in October--November, this site is as good as the best of studied summits in the world.
\end{abstract}



\section{Introduction}

The potential of a ground-based telescope is basically defined by the size $\beta$ of the image distorted by the earth's atmosphere. In classical astronomical observations, the telescope efficiency, in the sense of the detection limit, depends mainly on the $D/\beta$ ratio  \citep{Bowen1964,Shcheglov1980}. Other resources concerning improvement in the telescope optics quality and detector efficiency are already nearly exhausted. Further, the telescope yield relates to the same ratio being squared. Finally, the resolution-limited tasks are fully constrained by the image quality achievable at a given site.

The capabilities of ground based telescopes may be significantly boosted using various active and passive high angular resolution (HAR) techniques. Since they are quite complex to implement, it is important to assess and optimize their efficiency in advance. In turn, this requires one to be armed with statistically reliable and comprehensive information on atmospheric optical turbulence (OT)  including its vertical distribution, \citep[see for e.g.,][]{Vernin1991, Roddier1999, Wilson2003, Fuensalida2004}. Over the past two decades, such measurements are being conducted both at operating observatories \citep{Tokovinin2003MN,DaliAli2010,Wilson2009,Catala2013} and at potential extra-large-telescopes installation sites \citep{Shoeck2009,Vernin2011,Ramio2012,Thomas-Osip2008}.

On this basis, we set in 2006 the long-term program of monitoring of the OT vertical distribution along with other astroclimatic parameters at Mt.~Shatdzhatmaz in the Northern Caucasus, Russia, which is the site selected for the installation of new 2.5-m telescope of Sternberg Astronomical Institute (SAI).  Regular measurements were begun towards the end of 2007 and preliminary conclusions were presented in a paper two years later \citep{2010MNRAS}.

This study, apart from the explicit goal to optimize the performance of the 2.5-m telescope, is also more generally motivated towards a better understanding of the astroclimatic properties of the Northern Caucasus. Although site testing campaigns were conducted several times in this region \citep[full description is available in][]{Panchuk2011T}, their results are currently only of historical value.

The aim of this paper is to describe the main results of the six-year period of OT measurements (from November 15, 2007 to June 15, 2013) and to assess the stability of the main atmospheric parameters in  this time span. The second section presents the basic terms and definitions in use to describe OT, and the principles of the MASS and DIMM instruments widely used in OT monitoring throughout the world. Next, a brief description of the site and the campaign layout are outlined along with the main characteristics of the particular equipment used in the monitoring. The fourth section presents the main results of the OT parameter measurements and the analysis of their temporal behaviour. The last section focuses on the discussion of the determined properties of the OT at Mt.~Shatdzhatmaz. In particular, we focus attention on the suitability of this site for the application of HAR techniques.

In this article, we do not consider the results of the measurements of the main meteorological parameters, fraction of clear night sky, night-sky brightness, and atmospheric extinction as they are to be published in an upcoming paper.

\section{OT measurement methods}

\subsection{Basic terms}

The universally accepted treatment of atmospheric effects in astronomical observations is based on the theory of wave propagation through inhomogeneous media. OT is essentially the spatial fluctuations of the atmospheric refractive index that are adequately described by the Kolmogorov model with the only parameter $C_n^2$, the structural refractive-index coefficient. The model applicability at scales of 0.01--3~m  has been strongly validated. In typical conditions of astronomical observations, the effects of different atmospheric layers can be treated independently in accordance with the linear weak perturbations theory \citep{Tatarsky1967,Roddier1981}.

The sum effect of OT can be described by a number of quantities, all derived from the turbulence intensity $J$ (measured in $\mbox{m}^{1/3}$) defined as
\begin{equation}
J = \int C_n^2(h)\,\mathrm{d}h.
\label{eq:Jdef}
\end{equation}
Here, the integration is made over the relevant part of atmosphere. Unlike other integral parameters, the OT intensity $J$ does not depend on the wavelength $\lambda$ and is additive, which is convenient for the examination of its distribution along the line of sight.

Phase perturbations induced by OT are frequently parametrized by the Fried radius $r_0$ (atmospheric coherence radius) corresponding to the  phase structure function. The radius $r_0$ relates to the OT intensity as
\begin{equation}
r_0 = 0.185\,\lambda^{6/5}J^{-3/5}.
\label{eq:Jr0}
\end{equation}

The most popular characteristic among astronomers is the seeing, defined as the angular size of a long-exposure stellar image formed by an ideal large telescope. For the Kolmogorov model, it is given as
\begin{equation}
\beta = 0.98\frac{\lambda}{r_0}.
\label{eq:beta}
\end{equation}
Substituting (\ref{eq:Jr0}) in this equation we obtain (in arcseconds) $\beta = 2\cdot10^7 J^{3/5}$ for $\lambda = 500\mbox{ nm}$.

Considering the OT at various altitudes, we can observe its different features, generation mechanisms, and effects on astronomical observations. It is common to consider the OT over the first hundred meters as the ground layer (GL) turbulence, and in the first kilometer, it is referred to as the boundary layer turbulence. Higher layers are normally treated as free atmosphere, and the related OT intensities and corresponding seeing values are designated as $J_\mathrm{free}$ and $\beta_\mathrm{free}$, respectively.

An important complementary OT parameter is the isoplanatic angle \citep[see for e.g.,][]{Roddier1981} which, besides the intensity, depends on the OT altitude distribution:
\begin{equation}
\theta_0 = 0.314 \frac{r_0}{h^*},
\label{eq:thet0}
\end{equation}
where $h^*$ denotes the $C_n^2(h)$-weighted 5/3-power mean of the altitude $h$: ${h^*} = (\int C_n^2(h)h^{5/3}\,\mathrm{d}h)^{3/5}/J^{3/5}$.

The atmospheric time constant (coherence time) characterizes the rate of phase perturbation change due to wind translation. Its formal definition is given in \citep{Roddier1981} as follows:
\begin{equation}
\tau_0 = 0.314 \frac{r_0}{w^*},
\label{eq:tau0}
\end{equation}
where $w^*$ denotes the $C_n^2(h)$-weighted 5/3-power mean of the wind speed $w(h)$.

The parameters $\theta_0$ and $\tau_0$ determine the efficiency of the telescope's adaptive optics (AO). The first one describes its effective field of view over which the phase perturbations are mostly corrected, while the second one dictates the response time of the AO elements.

It is to be noted  that $\beta$, $\theta_0$, and $\tau_0$ all depend on the wavelength, and therefore, they are traditionally calculated for $\lambda = 500\mbox{ nm}$. In the following sections, we also use the same values.

\subsection{Principles of DIMM and MASS}

\label{sec:principe}

There is a considerable variety of astroclimatic instruments developed for OT measurement. All these instruments analyze  either the amplitude or phase perturbations of a light wave. Meanwhile, the most representative and homogeneous data series have been obtained with only two methods: DIMM \citep{Sarazin1990} and MASS \citep{2003SPIE}. In particular, these instruments have been intensively exploited in site testing programs of the following future extra-large telescopes: TMT \citep{Shoeck2009}, E-ELT \citep{Vernin2011}, and GMT \citep{Thomas-Osip2008}.

The DIMM measures the variance of the relative motion of two images of a single star that are formed in the feeding telescope focal plane by two subapertures $D$ ($\sim\!\!10$~cm diameter) separated by a distance $b$ ($\sim\!\!20$~cm). The theory of the method has been previously described in \citep{Sarazin1990,Martin1987,Tokovinin2002a}, wherein the measured variance is shown to relate linearly to the OT intensity $J$.

Being attractively simple and robust, the DIMM method underestimates the high-altitude turbulence for which the near-field approximation is not completely valid \citep{2007aMNRAS,2007bMNRAS}. In addition, finite exposure of its imaging camera smoothens the motion and lowers its variance \citep{Martin1987,Tokovinin2002a,2011bMNRAS}. Some other instrumental effects also need to be taken into account to avoid various systematical biases \citep{2007aMNRAS}. Nevertheless, the DIMM allows a fairly reliable assessment of the integral OT. This is necessary but not sufficient information.

The instrument suitable to measure the OT vertical distribution is the MASS. The underlying method is based on the simultaneous detection of stellar scintillation in four entrance apertures of different sizes. This results in four normal and six differential scintillation indices $s^2_j$ that are variances of the relative light flux fluctuations \citep{2003SPIE,2003MNRAS}. Although the MASS altitude resolution is far coarser than that of the SCIDAR \citep{Fuchs1998}, this instrument is  considerably more suitable for long-term monitoring and field applications at potential sites.

The scintillation indices $s^2_j$ relate to the OT distribution over the altitude in the following  manner:
\begin{equation}
s^2_j = \int_0^\infty C_n^2(z) Q_j(z)\,dz,\qquad j=1,\dots,10,
\label{eq:sc}
\end{equation}
where the values $Q_j(h)$ are precomputed weighting functions depending on the aperture size, source angular size, spectral light composition and adopted model of the refractive-index perturbations \citep{Tokovinin2002b,2003MNRAS}.

Solving the system of 10 linear algebraic equations,  derived from (\ref{eq:sc}) by a discretization over an adopted altitude grid, we obtain the discrete OT distribution \citep{2003MNRAS,2011ExA}. In practice, the weighting functions set $Q(z)$ makes this problem ill-posed, and therefore, the solution method is an important part of the data processing. A principal restriction of the MASS method is its insensitivity to GL turbulence, because all weighting functions $Q_j(0) = 0$.

The normal weighting functions $Q_j(h)$ are close to power-laws on altitude with exponent ranging from 5/6 to 2 depending on the aperture size. For the differential indices, functions $Q_j(h)$ are saturated after some specific altitudes \citep{Tokovinin2002b}. This allows fitting of the atmospheric turbulence moment $M_{\alpha}$ with power $\alpha$ by a linear combination of the measured indices $s_j^2$ \citep{2003MNRAS}.

Atmospheric moments can be used to directly compute integral OT parameters such as the free-atmosphere seeing  $\beta_\mathrm{free} \propto M_0$, isoplanatic angle $\theta_0 \propto M_{5/3}^{-3/5}$, and some other parameters. Evidently, the same parameters can be derived from the OT vertical profile, and their comparison can serve for profile restoration verification.

The scintillation indices measured with different exposure times allow the assessment of the atmospheric time constant $\tau_0$ \citep{Tokovinin2002b}. In the present work, a similar method that has been validated in \citep{2011A&A} is used. This estimation does not include the influence of the GL turbulence and thus is denoted as $\tau_\mathrm{free}$. The time constant $\tau_0$ for the whole atmosphere can be composed in the usual manner \citep[see for e.g.,][]{Travouillon2009a}:
\begin{equation}
\tau_0^{-5/3} = \tau_\mathrm{free}^{-5/3} + \tau_\mathrm{GL}^{-5/3},
\label{eq:tauGL}
\end{equation}
where $\tau_\mathrm{GL}$ denotes the time constant for the GL computed from the turbulence intensity $J_\mathrm{GL}$ and the  surface wind speed $w(0)$.

\section{Site testing campaign}

\subsection{General site description}

\label{sec:gen}

Caucasian astronomical observatory of the SAI is located atop  Mt.~Shatdzhatmaz (2127~m) in the Northern Caucasus region of Russia. This summit is part of the Skalisty ridge that runs parallel to the Main Caucasus Ridge some 50~km to the north. In this region, there are two other astronomical observatories. Located 100~km to the west is the 6-m telescope of the Special Astrophysical Observatory, situated at the northern branch of the main ridge. The Terskol observatory (with a 2-m telescope) is located 50~km to the south, lying on the immediate eastern slope of the Elbrus volcano. Thus, unlike the SAI site, both these observatories are situated near higher peaks.

\begin{figure}[h]
\centering
\includegraphics[height=9cm]{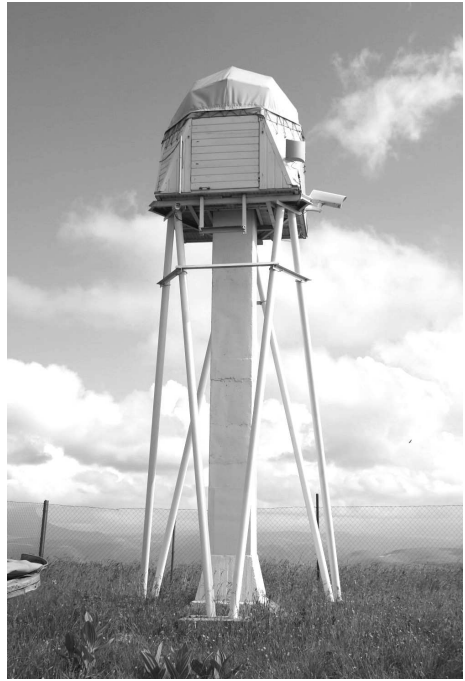}
\caption{Automatic site monitor at Mt.~Shatdzhatmaz as viewed from the north. \label{fig:view}}
\end{figure}

The 2.5-m telescope tower is erected 40~m away from the steep southeast slope of the mountain, at an  elevation of 2112~m. The telescope GPS coordinates are $+43^\circ44^{\prime}10^{\prime\prime}$~N,  $+42^\circ 40^{\prime}03^{\prime\prime}$~E. The astroclimatic monitor is located 45~m along the WWS direction from the telescope.

The climatic characteristics of the site are listed in \citep{2010MNRAS} based on our measurements  obtained over the period 2007--2009. The median annual temperature is $+1.8^{\circ}$C while its peak-to-peak amplitude on clear-sky nights varies from $-17.2^{\circ}$C to $+17.8^{\circ}$C. The surface is an alpine meadow slightly covered by snow in winter. This facilitates a mild day-to-night thermal contrast for clear skies, which is on average $-1.3^{\circ}$C between one hour before and one hour after sunset. The average relative humidity is 80--85\% in summer while in winter it is drier, $\sim\!\!65$\%. The median precipitable water vapor for observation nights is 7.75~mm \citep{Voziakova2012AstL}.

Night-time winds blow mostly from the west along the smooth-shaped ridge or from the southeast across a river valley with depth of several hundred meters. The median wind speed is $3.3$~m/s and $2.3$~m/s during day and night, respectively. Winds stronger than 9~m/s were recorded for only 2.2\% of clear  nights. Yet, storms with wind speeds up to 35~m/s have been recorded yearly.

\subsection{Measurements and statistics}

\label{sec:ameba}

In order to obtain the basic astroclimatic characteristics, namely, the statistics of the OT vertical distribution, clear night-sky time, and local meteo parameters such as the surface wind speed and direction, ambient temperatures, and humidity, we have developed an automatic seeing monitor (ASM). The details of the apparatus and functioning algorithm have been described in \citep{2010MNRAS, 2012ASIC}; thus, here we only recall its key features.

The MASS/DIMM instrument is attached to a 12-inch Meade RCX400 telescope installed atop a 5-m-high concrete pillar. The tube elevation is thus 6~m above the ground, which is typical for similar monitors at other observatories and slightly less than the altitude axis of the 2.5-m telescope (8~m).

The ASM was installed in the summer of 2007, and first trial measurements were performed in October. Regular observations under remote control began in November in parallel with the solving of various technical problems. The resolution of technical issues took several months, and therefore the monitor started to function in the fully automatic mode only in February 2008.

The informational structure of the ASM represents a typical distributed system. It involves three machines under the GNU/Linux OS and a set of digital devices connected to them. Software modules are started and run independently and communicate to each other using TCP/IP connections. This ensures global system stability and allows considerable simplification of the programs structure since a part of functioning is OS-driven \citep{2012ASIC}.

A special supervisor program is used to manage other software components using a dedicated simple command protocol. This program is restarted daily on a server suited for data archiving and ASM control from the Web. The system logic is described in \citep{2010MNRAS, 2012ASIC} in detail, which is summarized below in brief.

Observations are begun provided that 1) the Sun's altitude is lower than $-12^\circ$ (nautical night), 2) a clear sky is detected by a cloud sensor, 3) wind speed is less than 9~m/s, and 4) the ASM's power supply voltage is normal. Each observational session includes hardware initialization, target selection and pointing, measurement, and error handling. In case of a fatal error or worsening external conditions the session is shut down; the hardware is parked and powered off, and the enclosure is shut. Upon sunrise, the session is shut down and the supervisor program exits.

In March 2008, the main algorithm was extended to include atmospheric extinction measurements \citep{Voziakova2012AstL}, and in October~2009, the OT measurements in the evening and morning twilight times were also included \citep{Safonov2011arXiv}. These auxiliary measurements are not analyzed in this paper.

\begin{figure*}
\centering
\includegraphics[height=11cm,angle=-90]{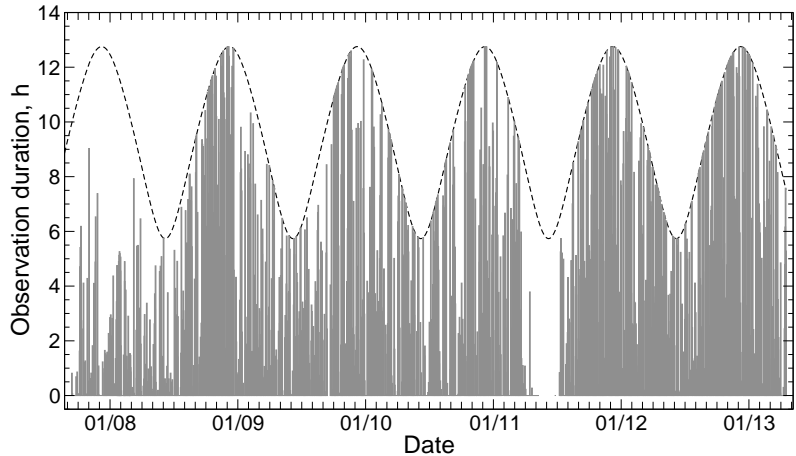}
\caption{Bar diagram of observation nights through the whole campaign period. The dashed line depicts  the nautical night duration.\label{fig:duty}}
\end{figure*}

Figure~\ref{fig:duty} shows the bar diagram of the observation duration for each calendar night throughout the campaign from November 15, 2007 to June 15, 2013. The total observation duration defined by the enclosure open/close time is 6669~h of which 776~h are due to twilight observations. The observed low data collection until mid-2008 is explained by the limited power supply in this period. The gap in the summer of 2011 is due to breakdown of the Meade telescope in this period.

For the whole campaign, more than $300\,000$ one-minute measurements were taken (total duration 5117~h). The overhead for auxiliary operations such as pointing, ($\sim\!\!15\,000$ pointings were made), centering, target verification, guiding loop adjustment, and background measurement ($\sim\!\!34\,000$ measurements) accounts for 18\% of the total observation time.

In terms of the total amount of obtained OT measurements, only the Pachon, Paranal and Armazones site testing campaigns in Chile \citep[see Table~1 in][]{2012AA} have obtained more data than our study. Data processing shows that the collected measurements have high quality and homogeneity.

\subsection{MASS/DIMM instrument}

\label{sec:instrum}

Our campaign, similar to other such campaigns \citep{Schoeck2009,Vernin2011,Thomas-Osip2008}, made use of a combined MASS/DIMM instrument. Its design, properties, and advantages are described in \citep{2007aMNRAS} in detail. The essential difference here is the conceptual co-processing of MASS and DIMM data during profile restoration, see section~\ref{sec:processing}.

In course of almost the whole campaign, the instrument with the serial number MD09 was in use, but starting in January 2013, the device MD41 was attached to the telescope instead. Both instruments are practically identical but differ in terms of the Fabry lens that builds an exit pupil in the plane of the mask defining the subaperture sizes. Consequently, the parameters used in the data processing also differ slightly.

\begin{table}[h]
\caption{Parameters of the instruments used in the 2007--2013 campaign. \label{tab:mainp}}
\bigskip
\centering
\begin{tabular}{l|rr}
\hline\hline
Device \ups    & MD09 & MD41 \\[4pt]
\hline
DIMM camera scale, arcsec/pixel\ups  & $0.612$ &  $0.668$ \\
DIMM base $b$, cm & 19.6 & 19.9 \\
Subaperture diameter $D$, cm  & 9.0 & 9.1 \\
Magnification $k$ in MASS channel & 16.3 & 15.2 \\[4pt]
\hline\hline
\end{tabular}
\end{table}

The most significant parameter of the DIMM is an image scale that allows for conversion of the differential motion at the CCD into the corresponding angular measure. The scale was determined both by using multiple stars ($\theta$~Ser with separation $22.3''$ and the Orion Trapezium) and by the sky rotation method that is more precise. The resulting scales for both instruments are listed in Table~\ref{tab:mainp}. The DIMM geometry ($b$ and $D$) as well as the MASS apertures sizes are defined by the physical mask dimensions and magnification factor $k$ of the ``telescope + instrument'' system. The latter was measured with $\sim\!\!2$\% precision routinely after each maintenance round of the instrument using the backward apertures illumination method \citep{2007aMNRAS}.

The validity of the resulting OT vertical profiles also depends on certain other instrumental parameters. The scintillation indices $s^2$ should be correctly reduced to account for the photon noise. This reduction involves the non-poissonity of photomultiplier and counter dead-time \citep{2007aMNRAS}, which were determined periodically using calibration measurements. The spectral response curve of the MASS detectors (including optics) was controlled by special observations of stars with a wide range of spectral types, similar to a previous approach \citep{2009AstL}. It is  noteworthy that starting from 8 January 2013, the 0.5-ms microexposure integration was used instead of previous  value of 1~ms.

The high-speed CCD camera Prosilica EC650 is used in the DIMM channel being connected to the computer using an IEEE1394 (Firewire) interface. An exposure time of 4~ms was used till December 2009 when it was reduced to 2.5~ms after it was proved that precision is not compromised thanks the low readout noise ($\sim\!\!10\ e^-$). With these exposures, the camera can acquire more than 200 frames per second using the windowed reading of $60\times100\mbox{ pixels}$ giving maximal statistical precision of the measured variance. Given that the gap between frames is very small, the image motion covariance is determined for finite exposure correction \citep{2011bMNRAS}.

The channel data acquisition is performed by the {\sl mass} and {\sl dimm} programs. The functional algorithm of the former is essentially identical to its predecessor, the {\sl Turbina} program, and it is described in \citep{2003MNRAS,2007bMNRAS}. The {\sl dimm} program has a similar principle of operation.

After each measurement cycle of 1-s duration, the required statistical moments of signals are computed and stored in the output file for subsequent processing with the {\sl atmos} program. The data accumulation is externally triggered synchronously in both channels of the instrument and takes 60~s.

\section{OT measurement results}

\subsection{Data processing}

\label{sec:processing}

As mentioned above, the joint processing of DIMM and MASS data formed one of essential features of the project. The MASS measurement accumulation period was taken as a reference for processing. The DIMM measurements that are categorized under this interval are included in the restoration of the OT profile. In case these data are absent (e.g., in conditions of extremely strong wind-caused tube vibrations), the OT profile is restored without GL.

\begin{figure*}
\centering
\begin{tabular}{cc}
\includegraphics[height=8.5cm]{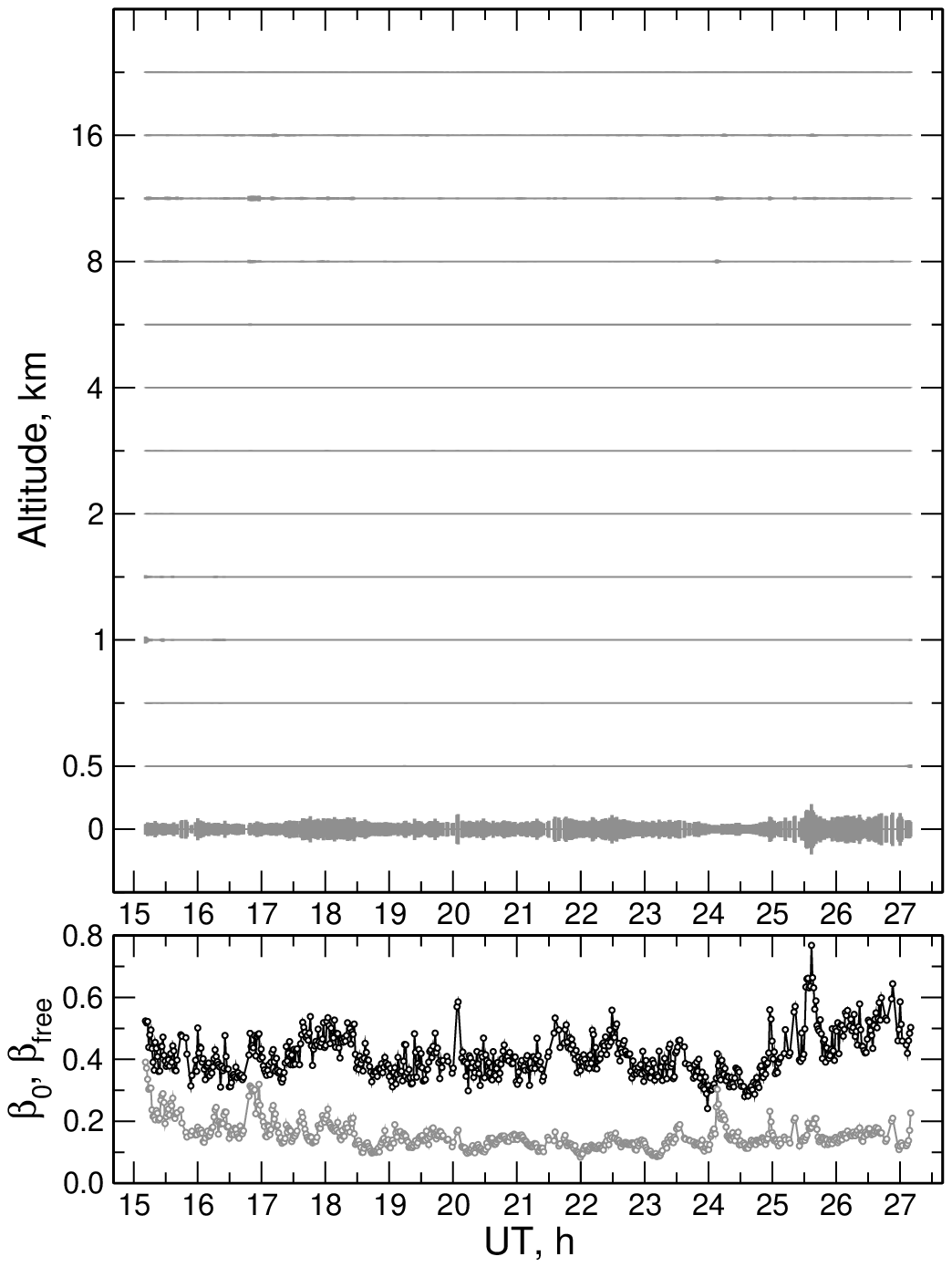} &
\includegraphics[height=8.5cm]{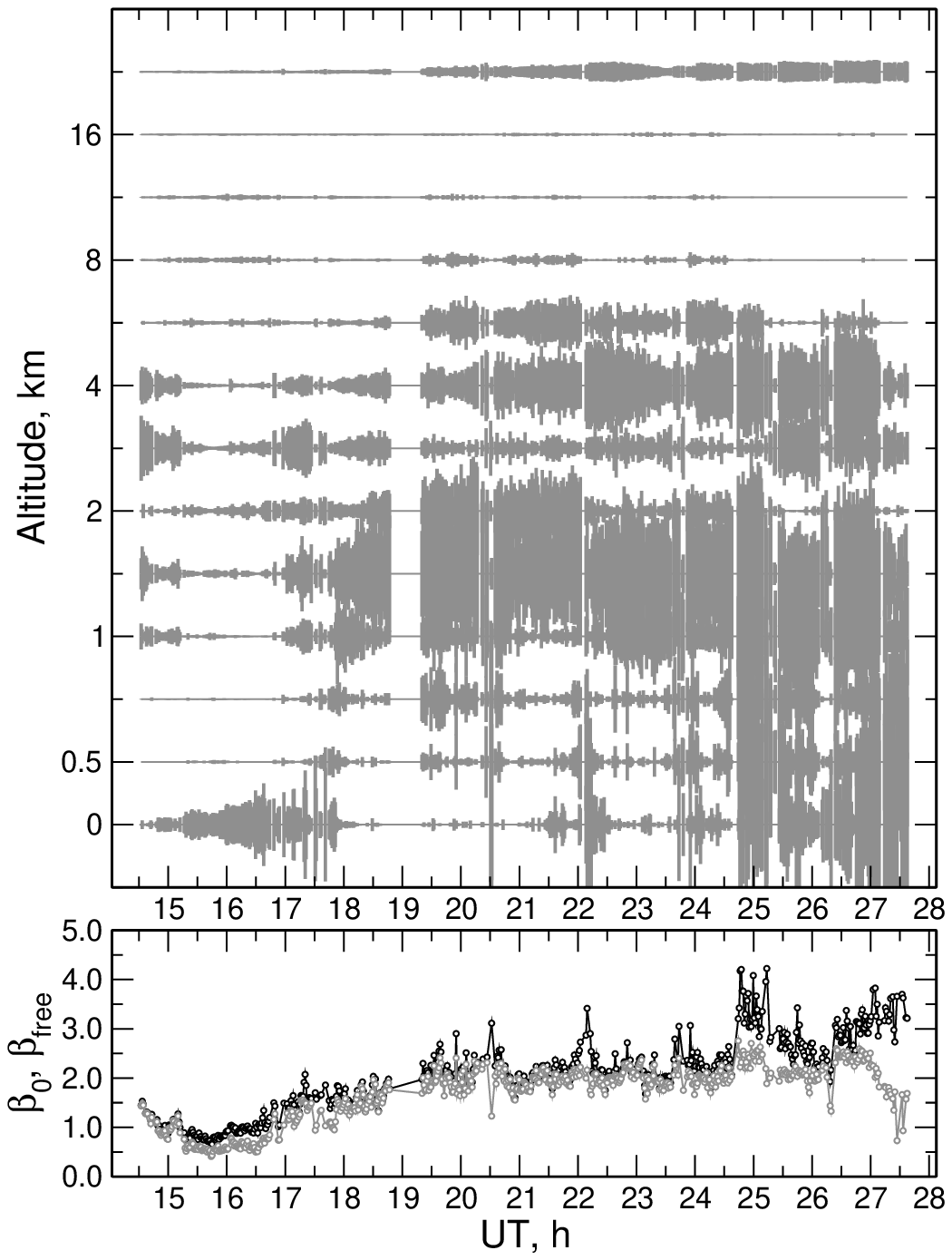} \\
\end{tabular}
\caption{Examples of OT profile restoration. On left: one of the best nights, October 20, 2009, on right: the strong-turbulence night, November 13, 2009. Lower panels show the behavior of the seeing $\beta_0$ (black dots) and free-atmosphere seeing $\beta_\mathrm{free}$ (gray dots). Medians of $\beta_0$ and  $\beta_\mathrm{free}$ for the first night: $0.40$\asec\ and  $0.145$\asec; for the second: $2.04$\asec\ and $1.96$\asec.
\label{fig:restor_example}}
\end{figure*}

The co-processing forced us to modify the OT restoration algorithm in the {\sl atmos} program. The non-linear minimization was replaced with the non-negative least squares technique (NNLS), and the equation system (\ref{eq:sc}) to be solved was extended; two relations for longitudinal and transversal variances $\sigma^2$ of image motion on the $C_n^2$ profile were added while the $h=0$ node was inserted in the altitude grid \citep{2011ExA,2010MNRAS}. To make this extension consistent, the DIMM related polychromatic weighting functions $W(z)$ are calculated in the {\sl atmos} \citep{MASS2010aT}.

Apart from the evident advantage of providing the GL turbulence intensity, this extended processing does also take into account the propagation effect in the DIMM measured values. Constraining the GL intensity to be non-negative additionally regularizes the solution. Based on previous considerations \citep{Tokovinin2002a}, we assume that the image motion detected by the DIMM camera is closely matched by the wavefront slope in terms of Zernike polynomials (z-tilt) rather than gravity centers (g-tilt).

The whole data set of the campaign was processed with the {\sl atmos-2.97.4} program, which includes the modified algorithm of the atmospheric time constant evaluation \citep{2011A&A}.

The filtering of data was performed according to the scheme described in \citep{2010MNRAS} with minor corrections. The OT twilight measurements were excluded as they required special processing. The cases of D-aperture counts lower than 100~counts/ms (clouds, escape of a star from the field diaphragm) or with a relative error of more than $0.05$ were discarded. Also we rejected the measurements made on the air masses larger than $1.3$ as these were obtained solely for extinction monitoring. Subsequent checking of the solution residuals revealed no need for any additional data cleanout. As a net result, we thus retained $284\,965$ vertical profiles and the corresponding integral atmospheric parameter sets.

\begin{figure}
\centering
\includegraphics[height=6.7cm]{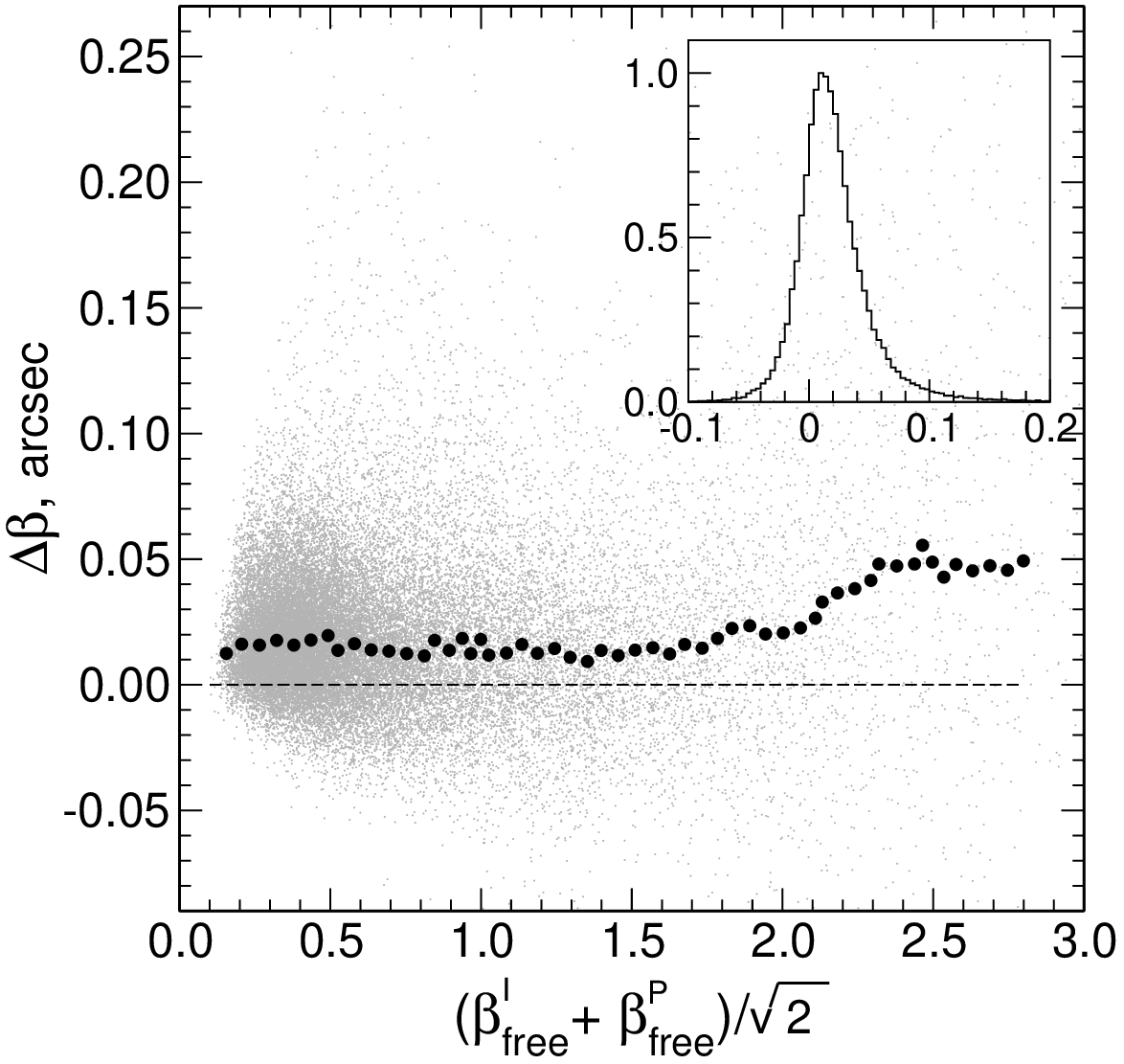}
\caption{Comparison of $\beta_\mathrm{free}$ computed from integral moments and from OT profiles. The dashed line represents the unity slope line while ordinates show the normal deviations $\Delta\beta = (\beta_\mathrm{free}^\mathrm{I} - \beta_\mathrm{free}^\mathrm{P})/\sqrt{2}$ from this line. The gray dots represent individual data points, while the black dots indicate medians of 501 point samples. The inset shows the differential distribution of $\Delta\beta$.
\label{fig:bcompar}}
\end{figure}

Following the practice of \citep{2010MNRAS}, the OT was restored on a logarithmic grid consisting of 13 altitudes: 0, 0.5, 0.71, 1, 1.41, 2, 2.82, 4, 5.66, 8, 11.3, 16, and 22.6~km. Examples of the profile restoration for two nights with different OT characteristics are presented in Fig.~\ref{fig:restor_example}. From the figure, we note that on the night of 20~October 2009, the GL turbulence was dominant while the free-atmosphere influence was extremely low. The example shown on the right presents a case of the weather change in November, although similar strong OT situations are more typical in January--March.

In order to verify the profile restoration, we compared the $\beta_\mathrm{free}$ estimates obtained by different methods (see section~\ref{sec:principe}). Interpreting the differences, it is necessary to take into account that the integral estimate of the free-atmosphere seeing $\beta^\mathrm{I}_\mathrm{free}$ accounts for part of the boundary layer turbulence, which otherwise is not considered in the $\beta^\mathrm{P}_\mathrm{free}$ values computed from the profile. In other words, some difference proportional to the ``extra turbulence'' in the layer from $0.4$ to $0.9$~km should be observed. In order to minimize this effect in comparison, we have selected the cases with minimal turbulence $J_1$ in the boundary layer (lower than its first quartile).

The result is shown in Fig.~\ref{fig:bcompar}, wherein the plot of $\beta^\mathrm{I}_\mathrm{free}$ versus $\beta^\mathrm{P}_\mathrm{free}$ is rotated by $45^\circ$ for clarity. Median points track the systematic difference that does not exceed $0.02$\asec, and only in cases of very strong free-atmosphere OT ($\beta_\mathrm{free} \gtrsim 1.5$\asec) the deviation is as high as $0.05$\asec. The individual points spread around the median relation does  not exceed $\pm 0.03$\asec, thereby being indicative of random errors introduced by the restoration technique.

\subsection{Seeing statistics}

\label{sec:seeing}

\begin{figure}
\centering
\includegraphics[height=6.7cm]{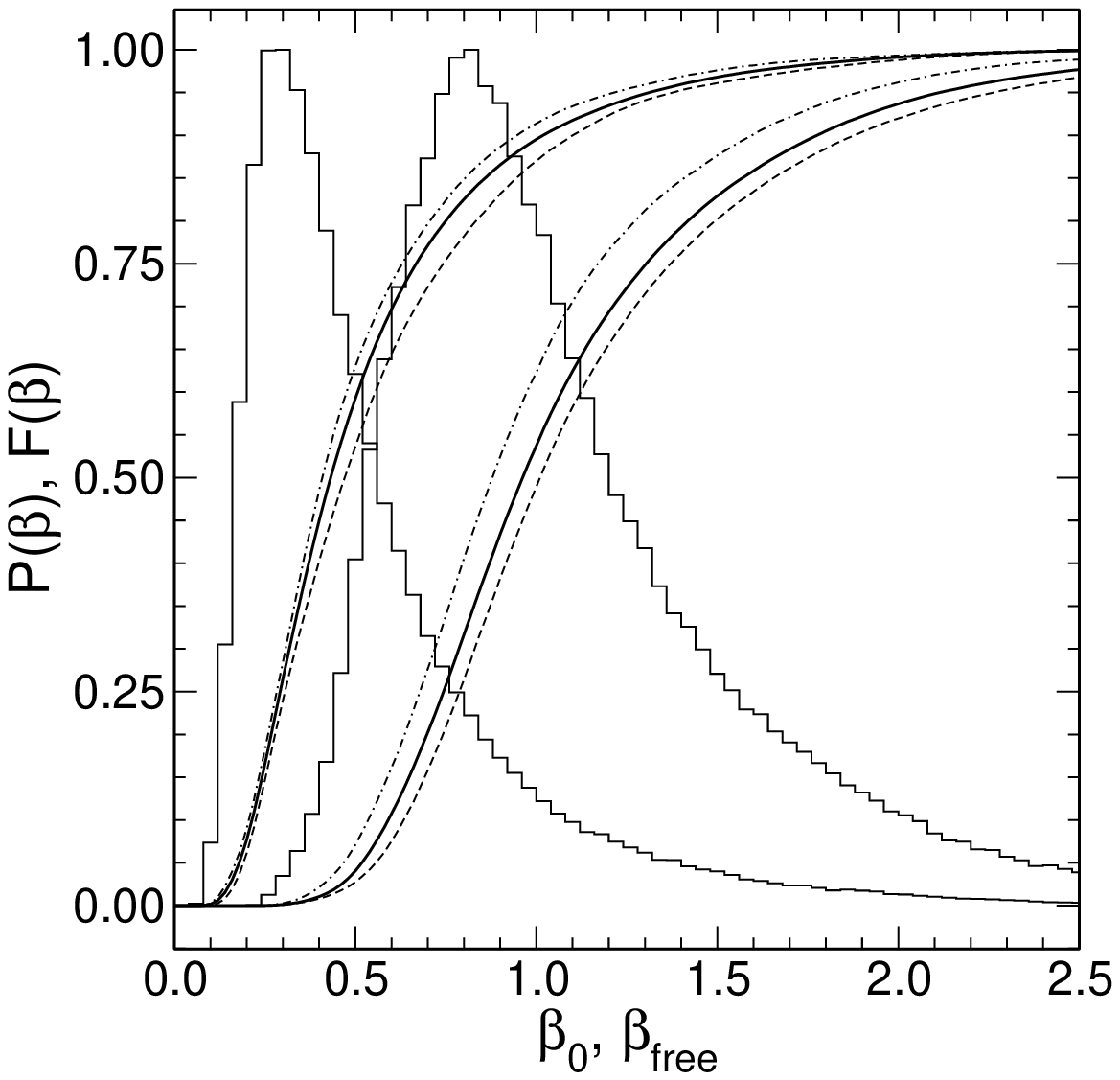}
\caption{ Cumulative distributions of $\beta_0$ (right thick curve) and $\beta_\mathrm{free}$ (left thick curve) for the whole campaign. Also shown are distributions for extreme seasons: $S10$ and $S13$ for full seeing and $S12$ and $S13$ for free atmosphere. Thin step lines show the differential distributions of $\beta_0$ and $\beta_\mathrm{free}$.
\label{fig:distr}}
\end{figure}

\begin{table*}
\caption{Medians $Q_2$ and quartiles $Q_1$ and $Q_3$ corresponding to integral seeing $\beta_0$, free atmosphere (1~km and above) seeing $\beta_\mathrm{free}$, and GL seeing $\beta_\mathrm{GL}$ for the seasons of the campaign (in arcseconds). $N$ indicates the data amount in a season.   \label{tab:seas_seeing}}
\bigskip
\centering
\begin{tabular}{l|rrr|rrr|rrr|r}
\hline\hline
Season \ups & \multicolumn{3}{c|}{$\beta_0$} &  \multicolumn{3}{c|}{$\beta_\mathrm{free}$} & \multicolumn{3}{c|}{$\beta_\mathrm{GL}$} & N \\
      & $Q_1$  & $Q_2$  & $Q_3$ &  $Q_1$ &  $Q_2$ & $Q_3$ & $Q_1$ &  $Q_2$ & $Q_3$ & \\[4pt]
\hline
all data   \ups  & 0.74 & 0.96 & 1.30 & 0.29 & 0.43 & 0.67 & 0.51 & 0.68 & 0.92 & 284\,965\\[3pt]
$S08$            & 0.73 & 0.94 & 1.36 & 0.31 & 0.45 & 0.69 & 0.48 & 0.64 & 0.88 &  13\,126\\
$S09$            & 0.76 & 0.96 & 1.25 & 0.31 & 0.44 & 0.65 & 0.50 & 0.67 & 0.89 &  55\,876\\
$S10$            & 0.68 & 0.88 & 1.17 & 0.27 & 0.42 & 0.66 & 0.45 & 0.60 & 0.80 &  52\,008\\
$S11$            & 0.75 & 0.98 & 1.39 & 0.28 & 0.42 & 0.68 & 0.54 & 0.72 & 0.95 &  41\,676\\
$S12$            & 0.74 & 0.98 & 1.33 & 0.30 & 0.47 & 0.74 & 0.48 & 0.65 & 0.88 &  53\,922\\
$S13$            & 0.79 & 1.01 & 1.37 & 0.28 & 0.41 & 0.63 & 0.60 & 0.77 & 1.01 &  68\,357\\[3pt]
\hline\hline
\end{tabular}
\end{table*}

\begin{figure*}
\centering
\begin{tabular}{cc}
\includegraphics[width=6.7cm]{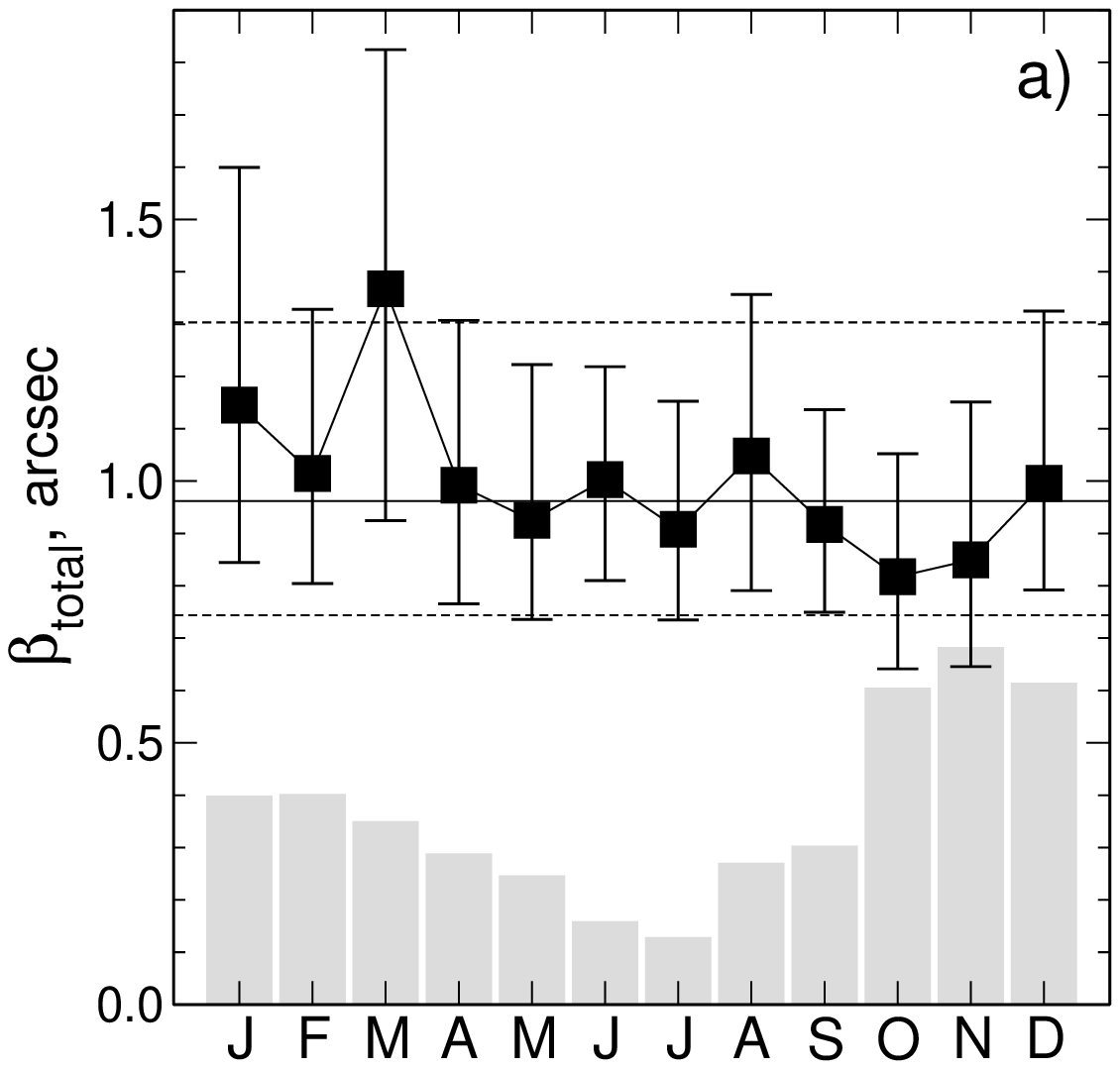}&
\includegraphics[width=6.7cm]{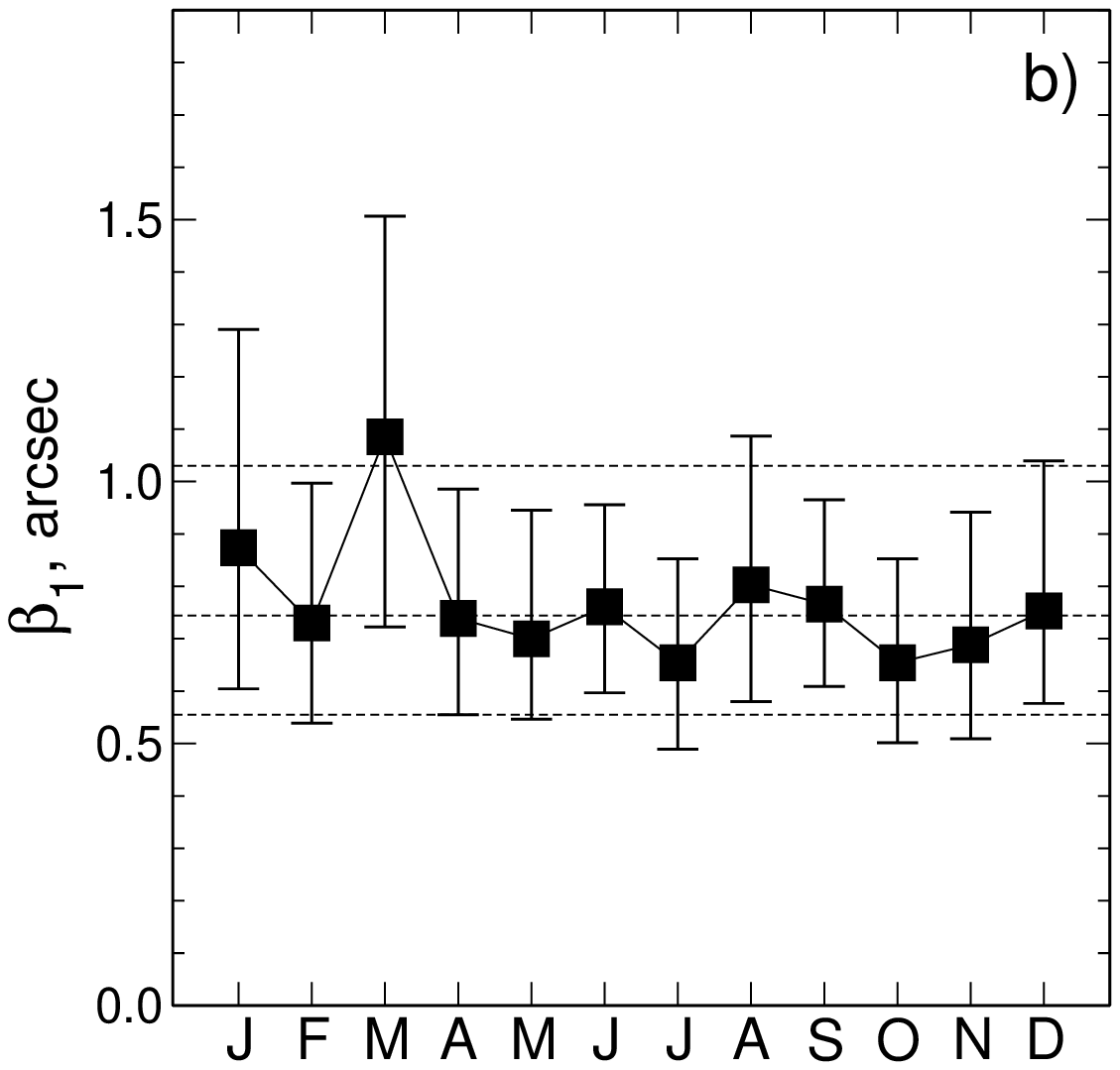}\\[-5pt]
\includegraphics[width=6.7cm]{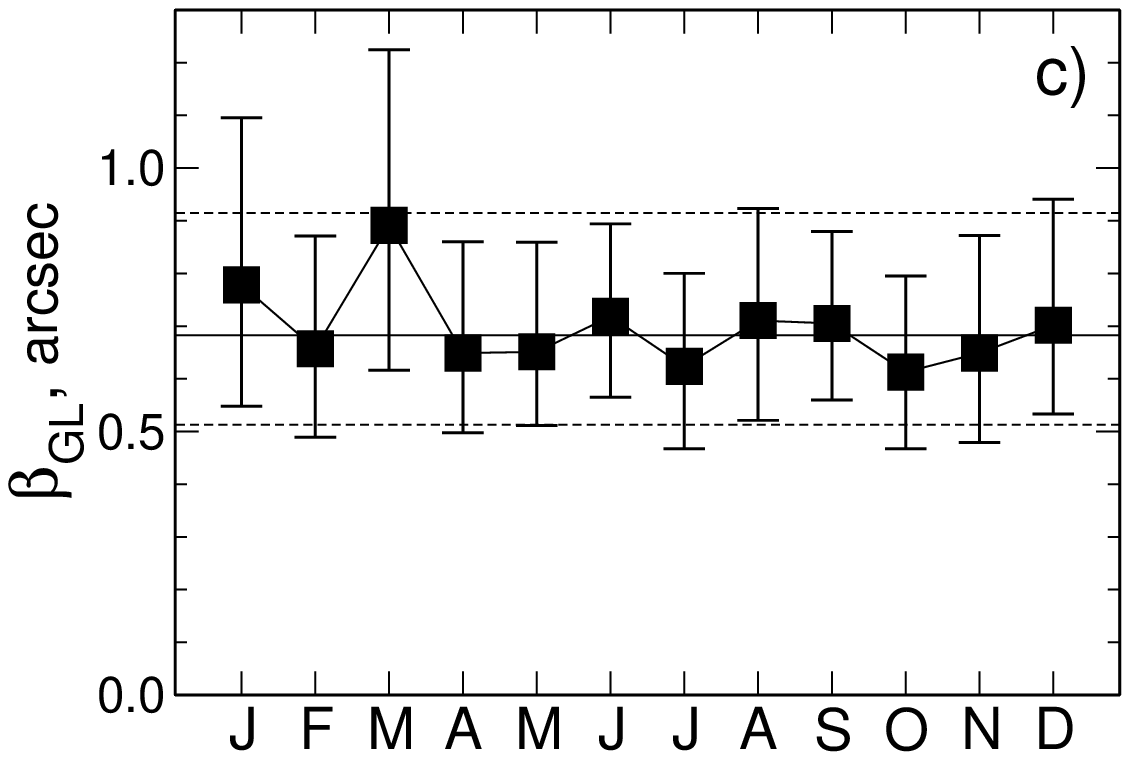}&
\includegraphics[width=6.7cm]{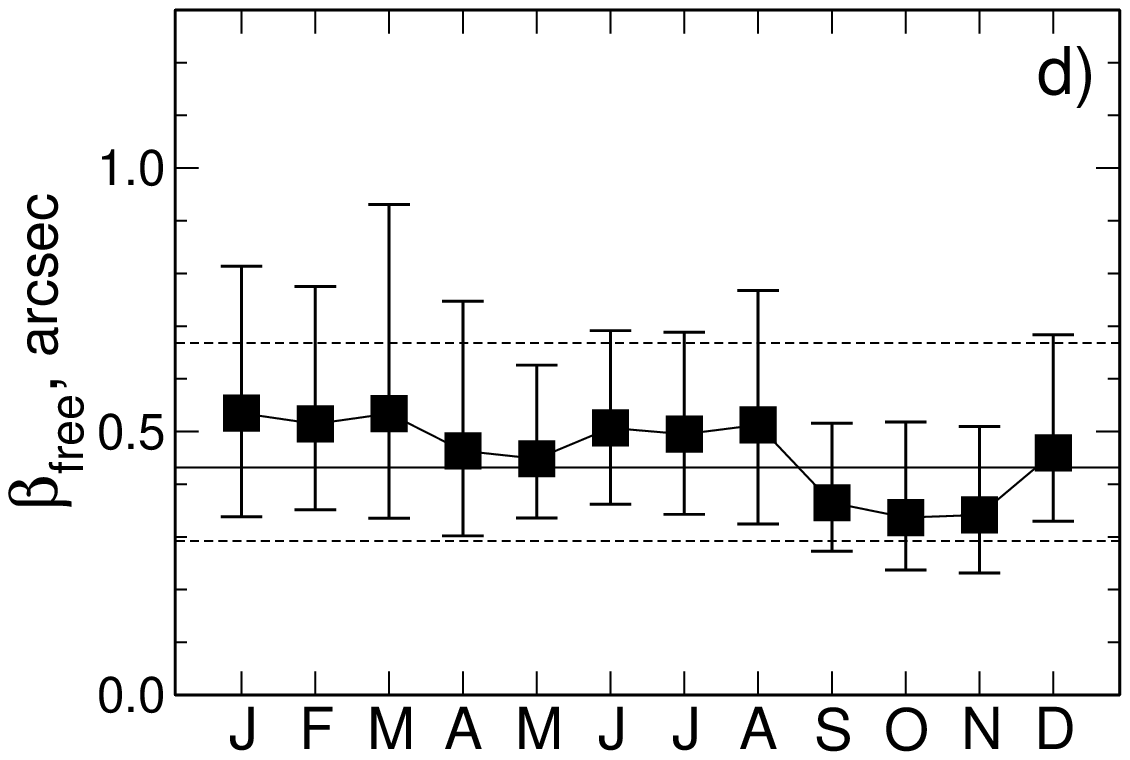}\\
\end{tabular}
\caption{The variability in OT folded onto a single year: a) whole atmosphere seeing $\beta_0$, b) boundary layer seeing $\beta_1$, c) seeing in the ground layer $\beta_\mathrm{GL}$, and d) free atmosphere seeing $\beta_\mathrm{free}$. Monthly medians are depicted by black squares, while the first and third quartiles are indicated by vertical segments. Horizontal solid and dashed lines depict the median and quartiles, respectively, for the whole data set. The gray histogram in the first panel shows the relative data set volume of each month. \label{fig:msee}}
\end{figure*}

For analysis of long-term OT variability, the data were categorized over ``annual seasons''. The ``new observational year'' was assigned to begin July 1 as the volume of summer-time data is relatively small and potential delays of annual phenomena do not significantly affect the statistics. These seasons are denoted as $SYY$, with $YY$ denoting the last two digits of the second calendar year of the season. For example, $S09$ refers to measurements made from July 1, 2008 to June 30, 2009. The statistical weights of the seasons are roughly similar except for the $S08$ being some four times less volumous in terms of data (see Table~\ref{tab:seas_seeing}).

Figure~\ref{fig:distr} shows the differential and cumulative distributions of the seeing $\beta_0$ and the free-atmosphere seeing $\beta_\mathrm{free}$. The distributions have a typical quasi log-normal shape with a moderate excess of high values. In the first approximation, we can consider the $\ln \beta$ and $\ln J$ values to be distributed normally. Differential distributions provide a most probable value (mode) of the seeing; $\beta_0$ has the mode value of $0.81$\asec\ and $\beta_\mathrm{free}$ has the mode value of $0.28$\asec. Curves for individual seasons do not differ in shape from the general one.

The statistical characteristics of the seeings $\beta_0$, $\beta_\mathrm{free}$, and the GL  seeing $\beta_\mathrm{GL}$ are listed in Table~\ref{tab:seas_seeing} for six seasons and for the whole campaign. It is evident that the turbulent atmosphere above Mt.~Shatdzhatmaz did not change principally during the campaign. Nevertheless, season-to-season variations can be observed; for e.g. in season $S10$, the OT intensity is about 14\% less than on average, while in $S13$, it is higher by 8\%.

Variations in the free atmosphere are also prominent but are not correlated with GL variations. Thus, in $S13$, the free atmosphere was even less turbulent than on average, which reveals the crucial role of both ground and boundary layers OT in the total seeing formation. This is natural for our site, where the OT is already known to be mostly driven by the GL turbulence whose median fraction is 65\% and the most probable one is 75\% \citep{2010MNRAS}.

Next, we compare the median seeing with those derived from the two-year measurements in \citep{2010MNRAS}: in 2010 the $\beta_0$ median was $0.93$\asec. The difference is insignificant and fully accountable by the sampling effects. The presented early results include a higher fraction of autumn measurements. To verify the comparison consistency, we estimated the median $\beta_0$ for the period considered in \citep{2010MNRAS} and we obtained a value of $0.92$\asec, which is in good agreement with the previous result.

In order to study the annual OT cycle, we folded all the estimations into one year and computed the statistics for each month. The obtained behaviors are shown in Fig.~\ref{fig:msee}. Naturally, the data volume for each month is dictated by the respective clear-sky fraction. The smallest data amount in July accounts for 8000 measurements (about 140 hours), which allows us to consider all these points as statistically significant.

The night-to-night variability in the seeing was studied by examining the quartiles of the $\beta_0$ distribution for each night having about 250 one-minute points on average, and in total accounting for 1132 nights. Nights containing less than 10 points were not considered. The cumulative  distribution of the night median $\beta_0$ is mimicked by the general distribution in Fig.~\ref{fig:distr} but is shifted to higher values by $\sim\!\!0.1$\asec. The median of this distribution is $1.06$\asec\ while the interquartile range $Q_3\!-\!Q_1$ amounts to $0.60$\asec, being slightly wider than that of the general distribution. From this fact, we can arrive at the following conclusion: the night-to-night seeing variations dominate over the annual and season-to-season variations.

The typical width of the $\beta_0$ distribution for one night was assessed via the interquartile range. Its median for all nights is $0.27$\asec. This means that the variability for a night is considerably smaller than the night-to-night variations. Here it is not accounted for an effect of proportionality of the interquartile range to the median value. A similar analysis of $\beta_\mathrm{free}$ reveals that nightly medians are distributed in a manner similar to the general cumulative distribution but with a median of $0.52$\asec. The typical interquartile range for a night is $0.18$\asec.

\subsection{Integral OT parameters}

\label{integrals}

In this section, we analyze the two most important integral atmospheric parameters: the isoplanatic angle $\theta_0$ and the atmospheric time constant $\tau_0$. These parameters are derived directly from atmospheric altitude moments without OT profile restoration (see section~\ref{sec:principe}).

\begin{figure*}
\centering
\begin{tabular}{cc}
\includegraphics[width=6.7cm]{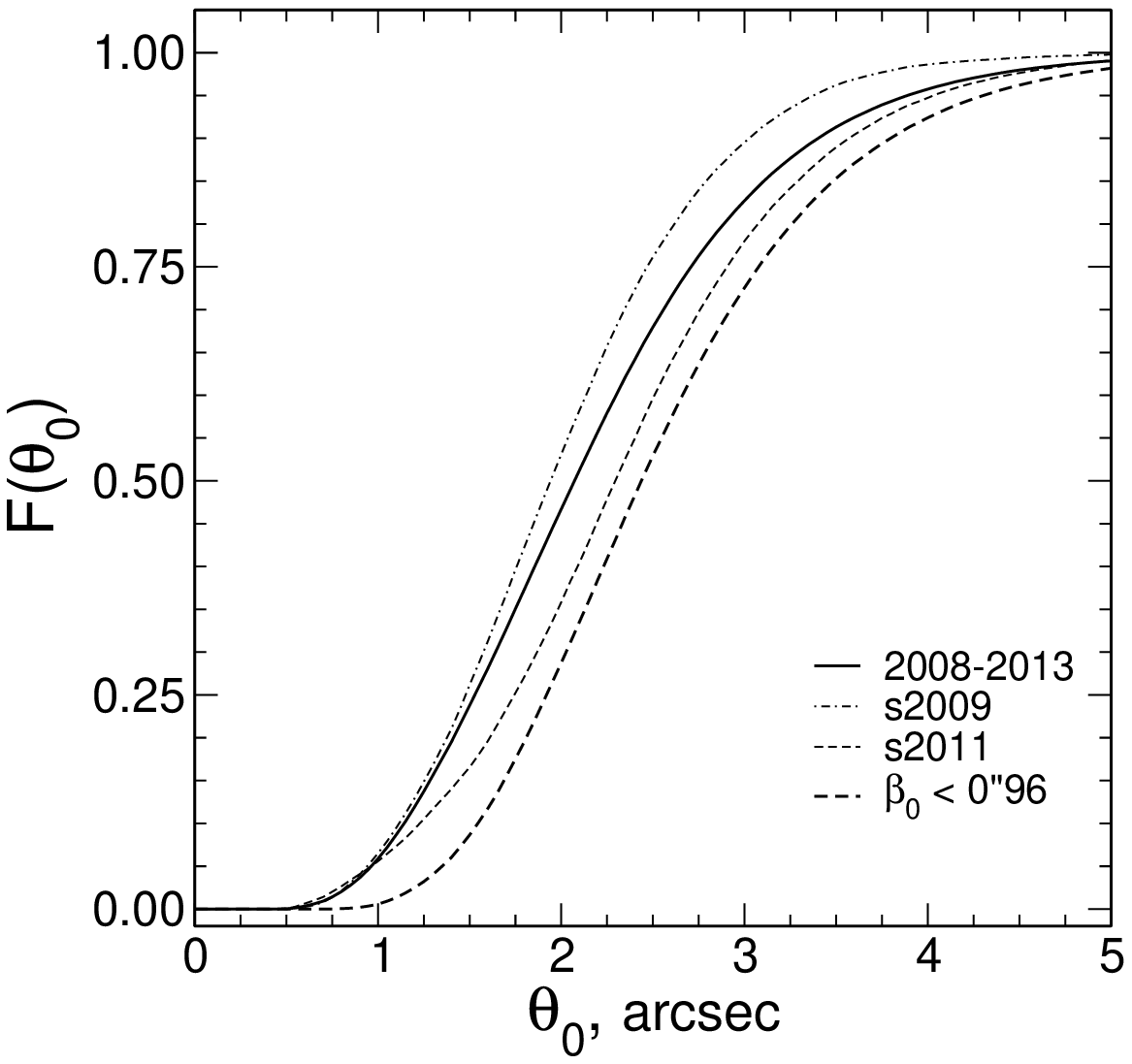}&
\includegraphics[width=6.7cm]{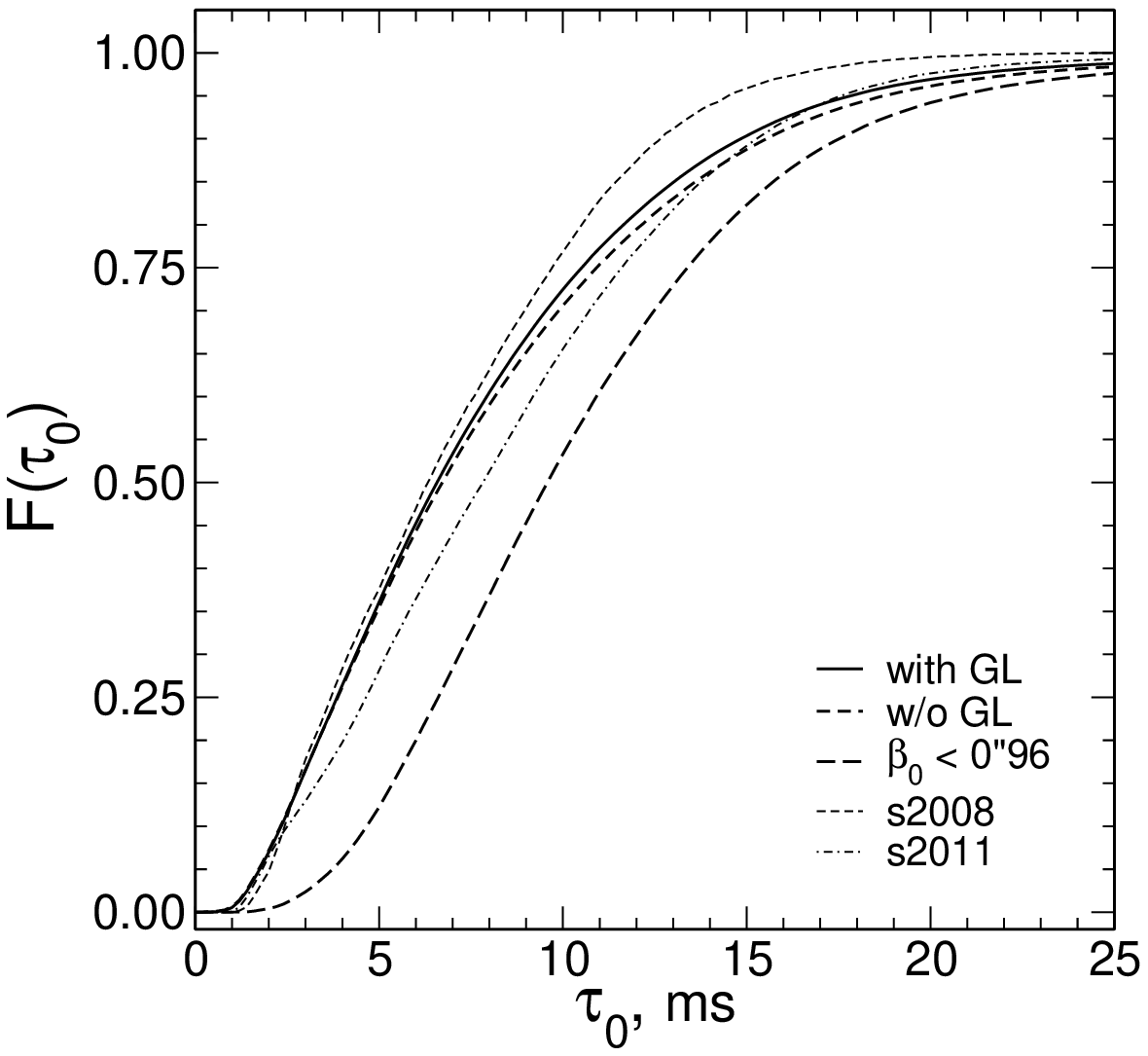}\\
\end{tabular}
\caption{Integral distribution of the isoplanatic angle  $\theta_0$ and atmospheric time constant $\tau_0$ for various data samples. Thick solid lines depict general distributions. The thick  short dashed line indicates the distribution of $\tau_\mathrm{free}$. The long dashed lines depict the distribution corresponding to ``good nights'' selection (as explained in the text). The best and worst seasons are shown by thin lines. \label{fig:dinteg}}
\end{figure*}

The $\theta_0$ values computed from OT profiles agree well with the integral estimates and may also be considered (see Fig.~13 in \citep{2010MNRAS}). Nevertheless, they have a lower precision due to the factor of the restoration process. In this light Fig.~\ref{fig:dinteg} shows the distribution of the integral $\theta_0$ estimates for the scope of the entire data and for the most standing out seasons $S09$ and $S11$.

The time constant $\tau_0$ is computed from the measured quantity $\tau_\mathrm{free}$ according to (\ref{eq:tauGL}). The influence of the GL time constant $\tau_\mathrm{GL}$ is computed from the GL turbulence intensity and the measured surface wind speed, and turns out to be very small $\sim\!\!1\!-\!2$\%. In case of the exceptionally calm upper atmosphere ($\tau_\mathrm{free} \gtrsim 20$~ms), the GL impact increases up to 10--20\%. The difference between the cumulative distributions of $\tau_0$ and $\tau_\mathrm{free}$ is seen in the right panel of Fig.~\ref{fig:dinteg}, and it is considerably lower than the corresponding difference between the ``extreme'' seasons $S08$ and $S11$.

The characteristic points of the statistical distributions of $\theta$ and $\tau$ obtained for the whole campaign and for particular seasons are listed in Table~\ref{tab:seas_integrals}. Here it is important to note an erratum; $\tau_0$ estimates presented in \citep{2010MNRAS} are invalid and unusable because of a software bug detected subsequently. This does not affect the time constant calculation algorithm used in this study.

\begin{table}
\caption{Medians $Q_2$ and quartiles $Q_1$ and $Q_3$ of the isoplanatic angle $\theta_0$ and atmospheric time constant $\tau_0$ for each season and for the whole data set. The last row corresponds to the measurements taken when the seeing was better than the median one. \label{tab:seas_integrals}}
\bigskip
\centering
\begin{tabular}{l|rrr|rrr}
\hline\hline
Season \ups & \multicolumn{3}{c|}{$\theta_0$, arcsec} &  \multicolumn{3}{c}{$\tau_0$, ms} \\
      & $Q_1$  & $Q_2$  & $Q_3$ &  $Q_1$ &  $Q_2$ & $Q_3$ \\[3pt]
\hline
all data \ups     & 1.53 & 2.07 & 2.71 & 3.86 &  6.57 & 10.51   \\[3pt]
$S08$             & 1.53 & 2.02 & 2.65 & 3.68 &  6.32 &  9.72   \\
$S09$             & 1.48 & 1.94 & 2.47 & 3.90 &  6.34 &  9.61   \\
$S10$             & 1.55 & 2.09 & 2.86 & 4.10 &  7.13 & 11.87   \\
$S11$             & 1.74 & 2.29 & 2.90 & 4.64 &  7.83 & 11.60   \\
$S12$             & 1.43 & 1.98 & 2.59 & 3.56 &  6.04 & 10.13   \\
$S13$             & 1.53 & 2.13 & 2.80 & 3.59 &  6.24 &  9.92   \\[3pt]
$\beta_0 < 0.96$  & 1.92 & 2.44 & 3.08 & 6.61 &  9.58 & 13.39   \\[3pt]
\hline\hline
\end{tabular}
\end{table}

It is seen that $\theta_0$ is relatively stable from season to season, having a full amplitude of 17\% only. The general median is nearly equal to the previously published value of $\theta_0 = 2.07$\asec. The maximal $\theta_0$ is observed in $S11$, which is remarkable for its calm upper atmosphere (see Table~\ref{tab:seas_seeing}). For comparison, at Mt.~Maidanak, an observatory located in Central Asia (latitude $\phi \approx +39^{\circ}$, altitude $\approx 2600$~m), the median isoplanatic angle equals $2.19$\asec\ and varies in the range of $1.92$\asec\ to $2.35$\asec\ from season to season \citep{2009AstL}.

Since the angle $\theta_0$ is mostly determined by turbulence conditions in the upper atmosphere, its value is only slightly affected by the intensity of the lower layers. Indeed, $\theta_0$ does not differ significantly from the isoplanatic angles obtained for other studied sites \citep{Skidmore2009,Vernin2011,Tokovinin2006MN}. Only Mauna Kea (Hawaii, $\phi\approx +20^\circ$) possesses a significantly wider isoplanatic angle with median $\theta_0 = 2.69$\asec, which may be due to its higher elevation (4200~m).

The general median of the time constant is $6.57$~ms, which exceeds the $\tau_0$ values of the five potential sites in the TMT program \citep{Travouillon2009a} and the four E-ELT sites \citep{Ramio2012}. The variation in $\tau_0$ is considerably stronger from season to season than those of the seeing values and isoplanatic angle. In our case, the amplitude of variation amounts to 30\%. The maximal $\tau_0$ value is observed in $S11$, wherein it approaches the value of $7.83$~ms.

We can suppose that the surface wind speeds used for $\tau_\mathrm{GL}$ calculation are underestimated due to low anemometer elevation. Checking this proposition we double the speeds (which is unlikely given that the wind speed at 1~km is only 4.3~m/s, see section~\ref{sec:disc}); however the GL influence increases up to 5\% only. Thus, in contrast to the case of Mauna Kea, wherein $\tau_\mathrm{GL} \approx \tau_\mathrm{free}$, \citep{Travouillon2009a}, the value of $\tau_0$ above Mt.~Shatdzhatmaz differs only slightly from $\tau_\mathrm{free}$ and is considerably large.

In order to assess the practical perspectives of the HAR methods, we examine the conditions yielding  good images. The distributions of $\theta_0$ and $\tau_0$ in cases of seeing being better than the median one, i.e. $\beta_0 < 0.96$\asec\ (``good images'' sample), are shown in Fig.~\ref{fig:dinteg} by means of the thick dashed line, and the corresponding characteristics are listed in Table~\ref{tab:seas_integrals}. It can be observed that the isoplanatic angle increases by $\sim\!\!17$\% while the time constant increases by almost 50\% approaching the value of $9.58$~ms.

\begin{figure*}
\centering
\begin{tabular}{cc}
\includegraphics[width=6.7cm]{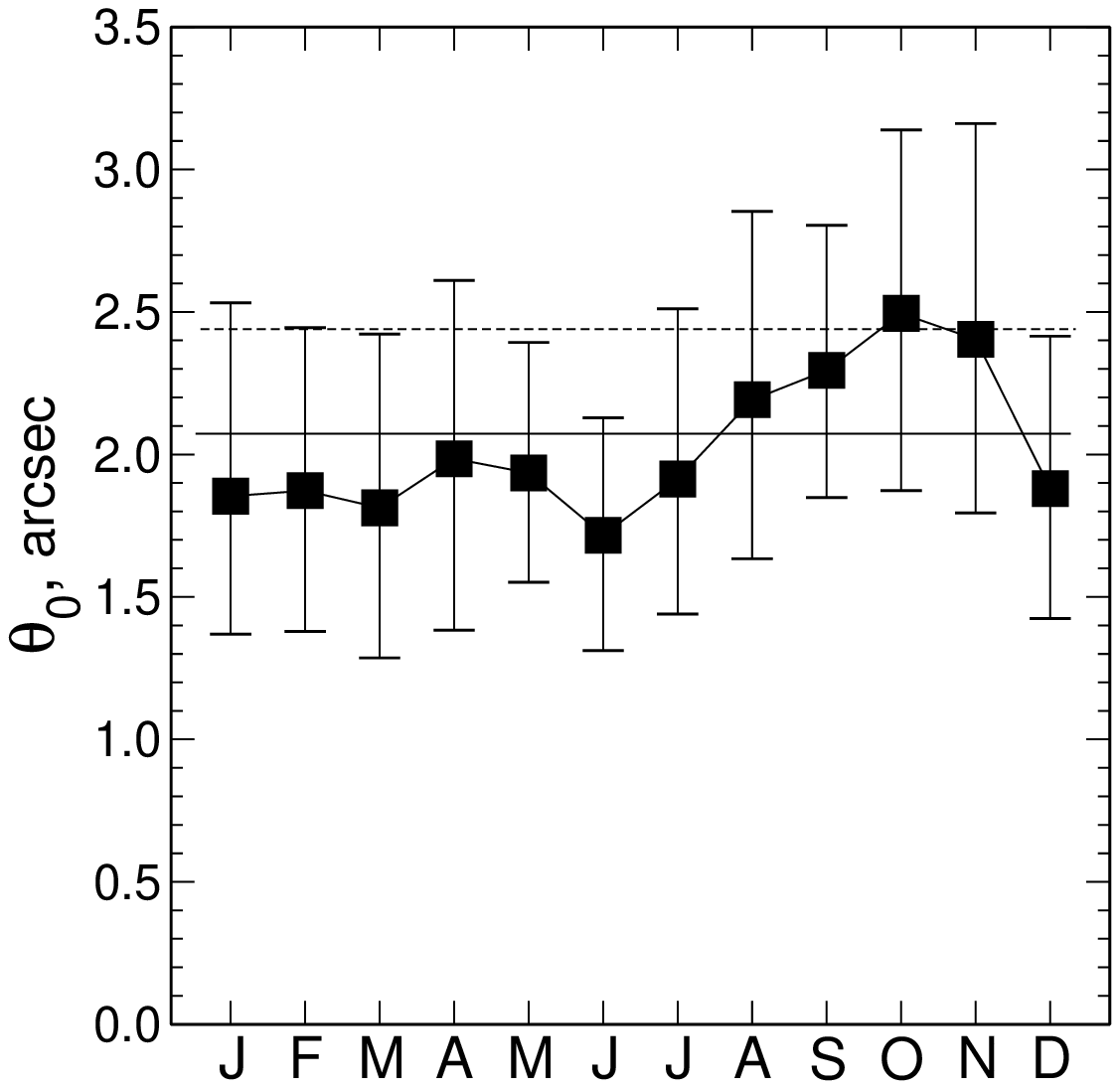}&
\includegraphics[width=6.7cm]{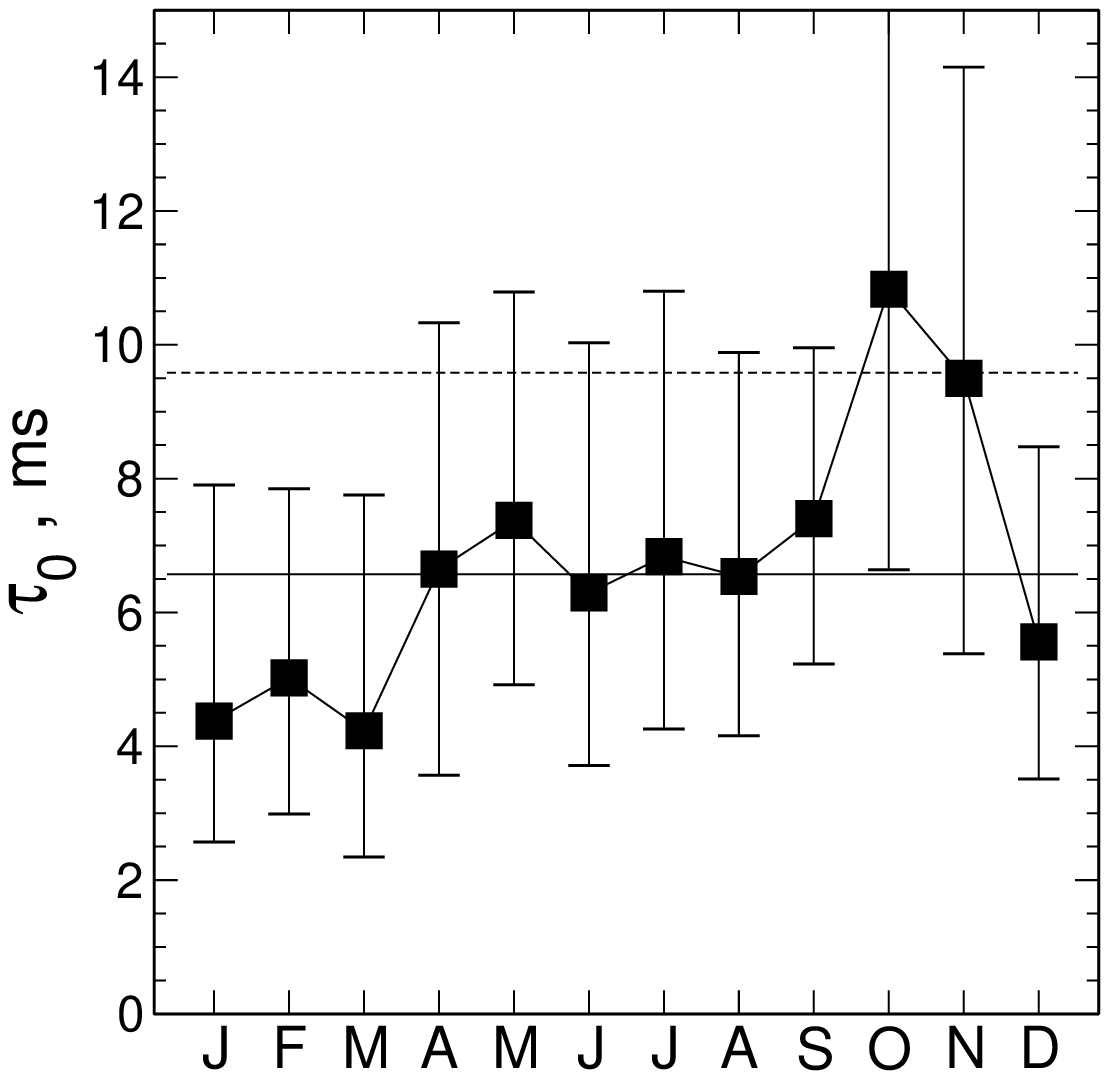} \\
\end{tabular}
\caption{Annual variation in the isoplanatic angle $\theta_0$ (left panel) and the atmospheric time constant $\tau_0$ (right panel). The squares indicate the monthly medians, while the vertical segments depict the interquartile range of the corresponding distributions. The thin horizontal lines depict the general median (solid) and the median for the selected ``good images'' (dashed). \label{fig:minteg}}
\end{figure*}

Both integral parameters vary significantly within a year since they are driven by an annual evolution of the upper atmosphere that is prominent at middle latitudes. These changes are illustrated in Fig.~\ref{fig:minteg} for $\theta_0$ on the left and $\tau_0$ on the right. It can be noted that the monthly $\theta_0$ median is greater than the general one in August to November only, approaching the value corresponding to the ``good images'' in October--November.

The atmospheric time constant is more sensitive to high-altitude winds. Hence, its annual variation has an amplitude of a factor of two; in winter time (December--March), this value is about 5~ms and it increases up to 10~ms by the end of autumn. During the rest of the time, the value of $\tau_0$ is close to its general median.

It is noteworthy that the magnitudes of OT parameters variations must not be extrapolated to observatories closer to the Equator. Annual variations in weather parameters are lesser at such locations and annual cycles of atmospheric parameters show less significant amplitude (nevertheless still noticeable \citep{2012AA}). Probably, the observed annual and season-to-season variations may explain the significant part of inter-site differences in the measured $\theta_0$.

The provided dependencies support the conclusion that the best observational period for HAR techniques at Mt.~Shatdzhatmaz is October--November, which period accounts for 28\% of clear skies.

\section{Discussion}

\label{sec:disc}

\subsection{Abnormal periods}

The annual behavior of the OT intensity (Fig.~\ref{fig:msee}) reveals two features. First, as already highlighted in \citep{2010MNRAS}, there is a prominent $\beta_0$ decrease in October and November compared to the seasonal median. The free-atmosphere seeing shows the same behavior in September--November, while for the GL turbulence such behavior is not so evident.

Surface winds measured directly show no significant variation in speed or in direction in autumn. This is as expected, since surface effects are unlikely to affect the free-atmosphere OT. It is more likely that the autumn here is remarkable for long-lasting anticyclones, which provide all the atmosphere up to the tropopause exceptionally calm.

The second feature concerns the ground and boundary OT. From Fig.~\ref{fig:msee}, we note that the intensity of this OT shows an increase in March. Such an effect, although considerably weaker, is also noticeable in January. A visual examination of the OT profiles suggests that March is characteristic of a high probability of the appearance of a strong turbulence in the lower atmosphere at $h<2$~km. Meanwhile, the median of surface winds speed is even lower in March than in July and October.

\begin{figure}
\centering
\includegraphics[width=6.7cm]{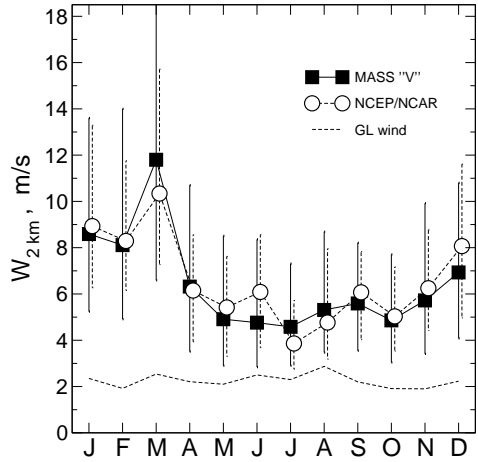}
\caption{Annual variances in the wind speed 2.5~km above the summit. Median values from MASS measurements are denoted by squares, while interquartile ranges are indicated by vertical segments. The circles depict NCEP/NCAR data for a pressure level of 600~mbar (data points are slightly shifted for clarity). The dashed thin line depicts the surface wind median speeds.
\label{fig:masswind}}
\end{figure}

To study the wind behavior in the free atmosphere, we reprocessed the MASS data using the {\sl atmos-}2.98.8  version that allows restoration of the wind profile via three different methods \citep{Kornilov2012SPIE}. Being folded into a single-year period, the results obtained by the method ``V'' (a direct minimization) are presented in Fig.~\ref{fig:masswind} as monthly medians and quartiles of the wind speed $w_\mathrm{MASS}$ at an altitude 2.5~km above the site. The surface wind speeds are also plotted. It is evident that in March the wind shear abruptly increases, corresponding to a vertical gradient 5~m/s per km. Such an increase in the gradient is probably the reason for the intensified OT generation.

For independent confirmation, we interpolated the NCEP/NCAR reanalysis data points for an altitude 2~km above the summit on the moments of our measurements. The medians and quartiles of  the corresponding monthly distributions are also shown in Fig.~\ref{fig:masswind}. The obtained median speeds correspond very well to the $w_\mathrm{MASS}$ values, and a sharp peak can be noticed for March against the background of the considerably smooth annual trend. A similar behavior of the wind speed from MASS data is seen for the 1~km layer but with a smaller amplitude.

The season $S13$ shows an increase in GL turbulence by $\sim\!\!20$\% (Table.~\ref{tab:seas_seeing}) and this made us consider as to whether this could be a consequence of the 2.5-m telescope tower erection. Further, during the year in question, large  containers with telescope parts were stored in the area between the tower and ASM, and a 500~W  spotlight was attached to the ASM under its platform. In order to analyze the possible influence of these obstacles on the local conditions, we divided the data set into two parts: before and after July 1, 2012.

The surface wind speed and direction statistics do not differ significantly for $S13$ and before. In any case, only 7\% of observations were made when the wind blew from the tower side, i.e. in the NE quadrant. The analysis of the OT intensity with respect to the wind direction does not indicate that the excess amount of $\beta_\mathrm{GL}$ depends on any particular wind direction. 

\begin{figure}
\centering
\includegraphics[width=6.7cm]{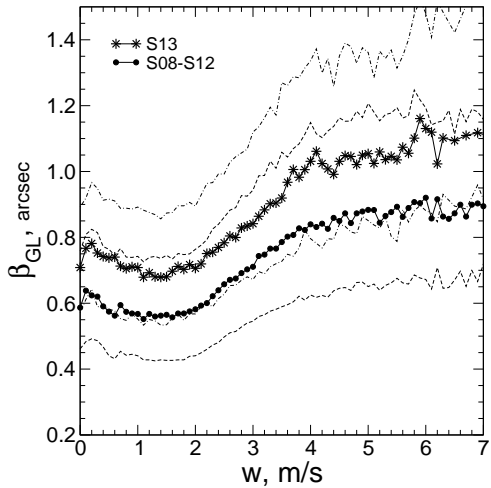}
\caption{Relation between GL turbulence and surface wind speed. Black dots indicate the medians for the data in the seasons $S08\!-\!S12$, and the asterisks depict the medians for season $S13$. The thin dashed and dot-dashed lines depict the $Q_1$ and $Q_3$ quartiles, respectively.\label{fig:wsee}}
\end{figure}

The $\beta_\mathrm{GL}$ relation with the wind speed also does not indicate that the origin of the observed effect may be local turbulence generated by the tower and containers. Otherwise, this relation should affect the shape of the dependence; under no-wind conditions, the airflow effect would not arise at all. Figure~\ref{fig:wsee} presents a contrasting picture: the dependence on the wind speed is shifted towards higher values of $\beta_\mathrm{GL}$. On the other hand, the effect of spotlight would be most promiment in no-wind conditions, when stream of warm air rises vertically and can, in principle, traverse the incoming beam.

In season $S13$, the median wind speed measured by MASS above the summit at $2.5$~km is the same as in the previous period. Meanwhile, the difference in wind speed between the surface and an altitude of $1$~km increased from $2.1$~m/s to $2.9$~m/s, i.e., by about 40\%. We tend to believe that this change is real and can cause strengthening of the GL turbulence, which is thus not related to the 2.5-m tower erection.

\subsection{Effect of incomplete clear nights}

\begin{figure}
\centering
\includegraphics[width=6.7cm]{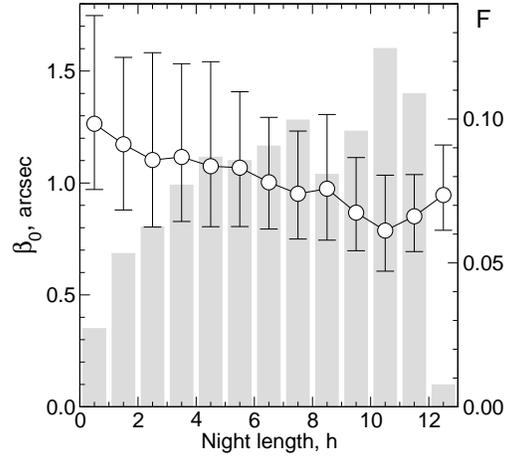}
\caption{Dependence of the median seeing $\beta_0$ on the observational night duration. Vertical segments indicate the interquartile range. The gray bars show the fraction $F$ of the measurements corresponding to each subset. \label{fig:nlength}}
\end{figure}

There is another observational bias in the seeing statistics related to the duration of an  observation night. To study this potential effect, we resampled the data into subsets according to the total duration of measurements in a given night with a 1-hour discretization. Figure~\ref{fig:nlength} shows the behavior of the quartiles of the respective $\beta_0$ distributions as a function of the length of the night.

For long nights (8 to 12 hours), the main selection factor is evidently the annual variation. Indeed, these nights correspond mainly to autumn months although some winter data are also present. Nevertheless, the median $\beta_0 = 0.79$\asec\ for nights with durations of 10 to 11~h is still lower than the best monthly median of $0.82$\asec\ in October.

The situation can be better understood by studying the leftmost part of Fig.~\ref{fig:nlength}. The medians of nights with duration shorter than 7~h are considerably higher than the general median (Table~\ref{tab:seas_seeing}), although short, incomplete nights are equally distributed over a year. This reflects the fact that incomplete nights mostly correspond to periods of weather transition.

\begin{table}
\caption{Medians of the integral seeing $\beta_0$ in samples arranged by duration of the observation nights. The number of data points $N$ in a sample, its fraction $F$ of the whole data set, and respective median in arcsec are listed. \label{tab:long_statistics}}
\bigskip
\centering
\begin{tabular}{l|ccc}
\hline\hline
Nights longer than \ups & $N$ & $F$ & $\beta_0$ \\[3pt]
\hline
1~hour\ups & 276976 & 0.97 & 0.955 \\
2~hour    & 261678 & 0.92 & 0.945 \\
3~hour     & 243679 & 0.86 & 0.937 \\
4~hour     & 222028 & 0.78 & 0.923 \\
5~hour     & 196652 & 0.69 & 0.910 \\
6~hour     & 172449 & 0.61 & 0.893 \\
7~hour     & 148933 & 0.52 & 0.877 \\
8~hour     & 118479 & 0.42 & 0.858 \\[3pt]
\hline\hline
\end{tabular}
\end{table}

This statistical dependence on the duration of observations may introduce a noticeable selection effect. While making OT measurements manually, operators subjectively avoid periods of unstable weather, missing short ``windows'' of a clear sky. The significance of this selection is reflected in Table~\ref{tab:long_statistics}. It is seen that by filtering out all nights shorter than 5~h, one can still obtain a representative data set of 70\% observations and an  ``improvement'' of $0.05$\asec\ in the seeing.

\subsection{Coherence \'etendue $G_0$}

Considering the optical efficiency as applied to AO techniques, it has been proposed \citep{Lloyd2004} to augment the traditional light grasp factor $A\Omega$ with the third coordinate $\tau_c c$. The resulting measure, the coherence \'etendue $G_0 = r_0^2\,\tau^{}_0\theta_0^2$, serves to assess the suitability of the turbulent atmosphere state for AO system operations. This factor was exploited by the ESO team to compare their four sites \citep{Ramio2012}.

It is straightforward to compute the coherence \'etendue from our data as well. The differential distribution of the $G_0$ appears exponential, similar to those of the sites studied in \citep{Ramio2012}, but in fact, it is a log-normal distribution with a very large variance. This is expectable because $\ln G_0 = 2\ln r_0 + 2\ln \theta_0 + \ln \tau_0$, and all the components of this sum are almost normally distributed. For our site, the $G_0$ median is $0.31$, which is a little less than that for the ORM site.

\begin{figure}[t]
\centering
\includegraphics[width=6.7cm]{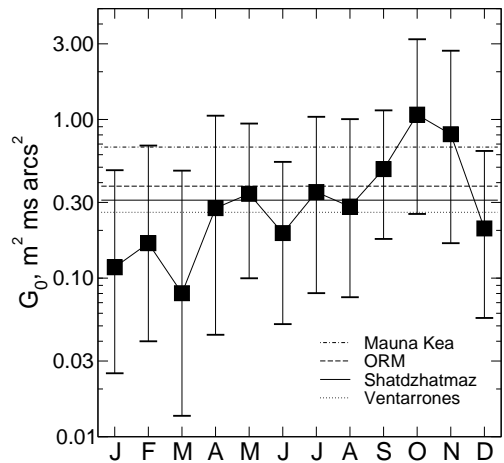}
\caption{Monthly medians of the coherence \'etendue $G_0$. Vertical segments indicate the interquartile range. The thin solid horizontal line depicts the general median for our data set. The estimation of $G_0$ for Mauna Kea (dash-dotted line) and the medians for ORM (dashed line), and Ventarrones (dotted line) taken from \citep{Ramio2012} are presented for comparison.
\label{fig:etendu}}
\end{figure}

While the $G_0$ seasonal medians vary between 0.25 and 0.45 (in seasons $S12$ and $S11$, respectively), its annual variation obtained by folding the data into a single year is  outstanding (Fig.~\ref{fig:etendu}), having a value of 0.1 in January to March and reaching up to 1.0 in October--November. One can envisage an abrupt change in the conditions for HAR methods application between November and December when $G_0$ drops from 0.8 to 0.2. Evidently, the continuous real-time monitoring of $G_0$ is a prerequisite for efficient time sharing between the traditional and HAR modes of observations.

Comparing this parameter between various sites, it should be noted that in general, the median of the product of variables may not be equal to the product of their medians. However, we confirmed that for the observed distributions of $\beta_0$, $\theta_0$, and $\tau_0$, the product of their medians is the same as the median of the true $G_0$ distribution, and it is equal to $0.31$.

Keeping in mind this caveat, we used data from \citep{Travouillon2009a,Skidmore2009} to compute the coherence \'etendue for sites examined in their studies. For three Chilean summits --- Tolar, Armazones, and Tolonchar located quite close to each other, the $G_0$ median values are practically identical ($\sim\!\!0.48$), although they lie at different altitudes of 2290~m, 3064~m, and 4480~m, respectively. San Pedro Martir situated in Baja California ($\phi \approx +31^\circ$), exhibits considerably lower \'etendue $G_0=0.28$. The maximal value of $G_0=0.67$ is observed at Mauna Kea, a completely different site, where high altitude and island geography are combined. Using data from \citep{2009AstL} and \citep{Ehgamberdiev2000}, we estimate $G_0$ at the Maidanak observatory to be as large as $0.60$.


%
%

\section{Conclusions}

The results of the campaign at Mt.~Shatdzhatmaz in 2007--2013 again confirm the high efficiency of OT monitoring in the automatic mode. In terms of the total amount of data collected on OT vertical distribution, this study is among the most representative ones. What is important is the location of the site in the Northern and Eastern hemispheres at mid-latitudes. As such, the closest analogues among actively tested observatories are those at San Pedro Martir and Mt.~Maidanak. 

We list the most significant results of this six-year monitoring. Here, we report on the median estimates of each OT parameter and view the season as a year-long period starting on July 1.

Over the whole period, the integral seeing $\beta_0$ is $0.96$\asec. For 25\% of time, $\beta_0 < 0.74$\asec\ while the most probable value is $0.81$\asec. The seeing in the free atmosphere (1~km and above) $\beta_{free}$ is $0.43$\asec, and the most probable one is $0.28$\asec.

From season to season, the changes in the seeing estimates have a relative range of $14$\% (which corresponds to $\sim\!\!23$\% in the OT intensity), both in the whole and in the free atmosphere. The best images are observed in October--November when the median $\beta_0$ is as low as $0.83$\asec. Strongest turbulence is encountered in March ($1.34$\asec) due to a significant wind shear in the lower atmosphere.

The isoplanatic angle is $2.07$\asec, which is typical for many observatories. Its estimates vary from season to season quite mildly, by $\sim\!\!10$\% of the general median. Peak-to-peak variations of monthly estimates approach 35\%, and in October, $\theta_0$ reaches $2.50$\asec.

The atmospheric time constant $\tau_0$ for the whole campaign period is $6.57$~ms. The season-to-season spread of the median estimates amounts up to 27\% while during a year $\tau_0$ changes between $4.2$~ms in March and $10.8$~ms in October.

A considerable variability in the basic OT parameters expressed both as annual and as season-to-season changes, confirms the need to conduct long-term site monitoring campaigns. Short-term calibration campaigns yield only rough estimates of the OT properties at a given location. This particularly applies to the atmospheric time constant whose variations are affected by several factors.

The coherence \'etendue factor $G_0$ \citep{Lloyd2004} that is considered a sensitive parameter in AO-related studies, exhibits the most considerable annual variation. In our case, it changes by an order of magnitude, reaching a value $\sim\!\!0.8$ in autumn. In this period, the Mt.~Shatdzhatmaz HAR capabilities are the same as those of Mauna Kea and Armazones.

\acknowledgments

Authors are grateful to all the colleagues from our institute and nearby Pulkovo observatory solar station who assisted in organizing and conducting the long-term measurement campaign at a location that initially has almost no infrastructure. We also thank our colleagues in the Astroclimatic Guild, especially to A.\,Tokovinin, M.\,Sarazin, and T.\,Travouillon, who inspired  us through their works.


\end{document}